\documentclass[final,3p,times]{elsarticle}

\usepackage{amssymb}
\usepackage{amsmath}

\usepackage{amsmath,amssymb,bm,graphicx,booktabs,microtype,xcolor,hyperref}
\hypersetup{colorlinks=true,citecolor=blue!45!black,linkcolor=blue!45!black,urlcolor=blue!45!black}

\newcommand{\pt}{p_{\mathrm{T}}}
\newcommand{\avpt}{\langle p_{\mathrm{T}}\rangle}
\newcommand{\Nch}{N_{\mathrm{ch}}}

\journal{Journal of Subatomic Particles and Cosmology}

\begin{document}

\begin{frontmatter}



\title{Theoretical Review on Bulk Properties and Light/Strange Hadron Production in Heavy-Ion Collisions}

\author[aaa,bbb]{Yuuka Kanakubo}
\affiliation[aaa]{organization={RIKEN iTHEMS},
             addressline={2-1 Hirosawa},
             city={Wako},
             postcode={351-0198},
             state={Saitama},
             country={Japan}}

 \affiliation[bbb]{organization={Lawrence Berkeley National Laboratory},
             addressline={1 Cyclotron Road},
             city={Berkeley},
             postcode={94720},
             state={CA},
             country={USA}}

\begin{abstract}
Since the discovery of large elliptic flow at RHIC in the early 2000s,
relativistic hydrodynamics has become a central framework for quantitative studies
of the quark--gluon plasma (QGP).
In parallel, increasingly sophisticated analyses have improved constraints on
the initial state and the transport properties of the QGP.
In this contribution, I overview the recent progress on bulk-property of QGP and light/strange hadron production.    
\end{abstract}


\end{frontmatter}



\section{Introduction}
\label{sec1}
Relativistic hydrodynamics has become the standard description of the dynamics of the soft sector, following the experimental discovery of the large elliptic flow at the RHIC. The azimuthal distribution of final state hadronic production is expanded as
\begin{equation}
 \frac{dN}{d\phi}\propto 1+2\sum_{n=1}^{\infty}v_n\cos[n(\phi-\Psi_n)] ,
\end{equation}
where $v_n$ is the Fourier coefficient, and $\Psi_n$ is the angle of the event plane. Especially, $v_2$ is referred to as elliptic flow. 
The experimentally observed large elliptic flow was successfully described by hydrodynamic simulations, where the final-state momentum anisotropy appears as a result of hydrodynamic response to the initial-state geometrical anisotropy.

There is a strong linear correlation between elliptic flow and the initial-state geometrical anisotropy $\epsilon_2$,
and the ``linear response'' is shown to be sensitive to the shear viscosity $\eta/s$ of QGP.

Over the last decade, one of the most significant progress in the bulk study is the application of Bayesian analysis to extract parameters from model-to-data comparisons~\cite{Bernhard:2019bmu,JETSCAPE:2020mzn,Nijs:2020roc}. These analyses have realized quantitative constraints on the temperature dependence of $\eta/s$. 
Although theoretical model uncertainty inheres in these results, the extracted values are shown to be of the same order as the KSS bound~\cite{Kovtun:2004de}.

Small collision systems have also become a major topic that has been intensively discussed in the soft sector, following the observation of several QGP-like signals in high-multiplicity $pp$ and $pA$ collisions \cite{ALICE:2024vzv}.
In particular, long-range azimuthal correlations resembling those observed
in heavy-ion collisions have been measured in these systems.
Their microscopic origin, however, remains under active discussion. For review, see for instance, Ref.~\cite{Schenke:2021mxx}. 

\section{Light-ion collisions}
Light-ion collisions provide a controlled system-size scan of such signals.
The interpretation of the QGP signals is well-established in heavy-ion collisions, while the interpretation in small collision systems remain less settled. 
The oxygen-oxygen ($OO$) collisions are expected to lie between these two systems in terms of their multiplicity~\cite{Huss:2020dwe}. The 2025--2026 oxygen programs at the RHIC and LHC therefore provide an important opportunity to investigate the systematic collision system size scan.

\subsection{Nuclear structure: $\alpha$-cluster}
Another important physics goal is to probe the nuclear structure of $^{16}$O, which connects the high-energy and low-energy nuclear physics.
Recent ab initio calculations predict nontrivial many-body correlations in \(^{16}\)O, which may generate cluster-like spatial structures ~\cite{Freer:2017gip}.
Since momentum distribution of final-state hadrons in relativistic heavy-ion collisions reflects information on the initial nuclear geometry, $OO$ collisions provide a new opportunity to investigate such nuclear structures experimentally~\cite{Giacalone:2024luz,Zhang:2024vkh}.

Results from the SMASH--vHLLE hybrid framework show that the elliptic flow from $OO$ collisions is larger compared to the pure-hadronic transport model (SMASH) or the string fragmentation model (Angantyr) ~\cite{Prasad:2024ahm,Constantin:2025ova}.
Overall, the preliminary experimental results reported from the CMS collaboration favor the description with hydrodynamics~\cite{ABaty:SQM2026}. 
Within these particular comparisons, however, the difference between configurations with $\alpha$-cluster and Woods--Saxon model is too small and not robust enough against the uncertainty originating from other stages.

\subsection{Small collision systems}
The description of the measured elliptic flow by hydrodynamic calculations
provides support for final-state collectivity in light-ion collisions,
although collective flow alone does not probe QGP formation.

In a deconfined hot medium, strange (anti-)quarks can be thermally produced and may approach chemical equilibrium during
evolution through sufficient secondary interactions.
As the collision-system size or multiplicity increases, the probability of producing strange quarks and antiquarks also increases.
Therefore, in the intermediate-sized collision systems, one can expect that the equilibrium may happens partially.

The core--corona models, a two-component model of QGP fluids and non-equilibrium partons, successfully describe the multiplicity dependence of strange-hadron production~\cite{Kanakubo:2019ogh,Kanakubo:2021qcw,Werner:2023jps}.
Within these models, the strangeness production ratios exhibit an approximate scaling with the charged-particle multiplicity across different collision systems. For $OO$ collisions at 20–30\% centrality, the corresponding multiplicity at midrapidity is approximately $dN_{\rm ch}/d\eta \sim 60$--$80$~\cite{SPucillo:SQM2026}. At the comparable multiplicity range, the core--corona models predict that roughly $70\%$ of the final-state particles originate from the QGP. This strongly supports the interpretation of the observables based on hydrodynamic models. At the same time, however, the sizable contributions of non-equilibrium component raise a question of how quantitatively applicable a hydrodynamic description is in $OO$ collisions.

\section{Initial-state modeling}
Hydrodynamic models are often initialized with parametrized entropy or energy densities, whose normalization is tuned to reproduce the experimentally observed $dN_{\rm ch}/d\eta$. Despite its simplicity and uncertainty, the studies with Bayesian analysis constrain the transport properties of QGP~\cite{Bernhard:2019bmu,JETSCAPE:2020mzn,Nijs:2020roc}.
However, microscopic interpretation of the initial state is clearly missing.
Therefore, the next step in bulk study would be to connect the obtained constraints by Bayesian parameter estimation and microscopic models.

Rapidity-dependent observables have been demonstrating the importance of longitudinal dynamics, represented by, for instance, longitudinal flow decorrelations and longitudinal dynamics of conserved-charge. 

For the recent development of 3-D initial state models, see for instance, Ref.~\cite{Carzon:2019qja,Garcia-Montero:2025bpn,Werthmann:2025ueu,Ross:2025qxr}. 
One of the challenge that can be addressed by these models is the description of initial conserved-charge distribution, especially, toward lower collision energies where baryon stopping becomes relevant.

Another challenge of initial-state modeling is to describe particle production consistently over different energy scales. In relativistic heavy-ion collisions, bundles of high-energy particles, ``jets'',
are produced from hard scatterings during the early stage of the collision. These hard partons traverse and interact with the evolving QGP. This interaction leads to the jet quenching, and carries information of QGP. Therefore, a unified theoretical description of the initial state production of both soft and hard partons is expected to have an important role on the jet quenching analysis.

The MC-EKRT framework ~\cite{Kuha:2024kmq,Hirvonen:2024zne} provides a microscopic description of the initial state with perturbative-QCD minijet production supplemented by a saturation conjecture. The incoming partons are sampled from spatially dependent nuclear parton distribution functions, and the produced partons are subject to a MC-EKRT saturation criterion, event-by-event energy-momentum and valence-quark number conservation. This provides a full three-dimensional
initial condition for subsequent hydrodynamic evolution. This study initializes hydrodynamics without the conventional normalization to reduce uncertainty in the initial state. 
Reproduction of $dN_{\rm ch}/d\eta$ becomes, therefore, a non-trivial task. The questions is whether a pQCD-based initial state with a limited number of tunable parameters can reproduce the observed multiplicity while respecting the energy-momentum conservation of incoming nucleus.

Using the MC-EKRT initial conditions followed by a viscous hydrodynamics model, a good description of the rapidity dependence of charged-particle multiplicity is obtained at both RHIC and LHC energies.
This provides a promising direction toward reducing the uncertainty in the initial states.

\section{New observables}
New observables have been providing the opportunity to investigate different aspects of the QGP in relativistic nuclear collisions, at the same time, challenging existing theoretical models to improve their description by introducing sophisticated physics.
Two observables that received particular attention at the conference are discussed below.

\subsection{Measurement of speed of sound}
At vanishing baryon chemical potential, lattice QCD determines the equilibrium equation of state (EoS). According to thermodynamics, the speed-of-sound $c_s$ is defined as:
\begin{equation}
 \label{eq:cs}
    c_s^2
    = \frac{dP}{d\epsilon}
    = \frac{d\ln T}{d\ln s},
\end{equation}
where $P, \epsilon, s, T$ are pressure, energy density, entropy density, and temperature, respectively.
It was proposed that the speed-of-sound of QCD equation of state can be inferred from observables by,
\begin{equation}
 c_{s,\mathrm{eff}}^2\simeq
 \frac{d\ln\avpt}{d\ln \Nch}
 \label{eq:csproxy}
\end{equation}
where $\left\langle p_T \right\rangle$ is a mean transverse momentum and $N_{\rm ch}$ is the number of charged particle multiplicity~\cite{Gardim:2019xjs}.
The CMS collaboration extracted this effective speed-of-sound, $c_{s,  \rm eff}$, from the slope of $\avpt$ versus $N_{\rm ch}$ in ultra-central Pb+Pb collisions and found consistency with lattice-QCD EoS~\cite{CMS:2024sgx}. Related analysis and theory studies extended the method across different collision energies and quantified corrections from non-equilibrium fluctuations~\cite{Mu:2025gtr,Alqahtani:2026edr}.

This has been high-lighted as a ``direct'' evidence for a deconfined QCD phase, and a stringent constraint on EoS of QCD matter. However, at the same time, several concerns have also been pointed out on this measurement. 
In Ref.~\cite{Gavassino:2025bts}, it is pointed out that the identification of $c_s^2$ from Eq.~\eqref{eq:cs} and $c_{s,\rm eff}^2$ from Eq.~\eqref{eq:csproxy} relies on several nontrivial assumptions.
First, one of the assumptions $\avpt = C_{\rm coeff} T$ with $C_{\rm coeff} \sim 3$ utilizes the results obtained from hydrodynamic simulations which already use information of the lattice QCD EoS.
Second, the relation $\avpt \propto T$ can be, more precisely, 
$\avpt \propto E/S$ at a fixed effective volume.
Moreover, the mapping from the underlying thermodynamic quantities to the measured $\avpt$ or $\Nch$ is affected by freeze-out, hadronic afterburner and kinematic cuts. 

The observable, $c_{s, \rm eff}$, still reflects a certain information of thermodynamic variables of QCD matter, however, describing this result as a direct measurement of the lattice QCD EoS may require some qualification.
Bayesian parameter estimation from model-to-data comparisons should be considered as a supplemental method to extract QCD EoS~\cite{Pratt:2015zsa} at the same time.

\subsection{$v_0(p_T)$ -- Radial flow fluctuation}

Radial flow fluctuation, $v_0(p_T)$, is defined from the covariance between the normalized spectrum in a momentum bin and a reference mean transverse momentum,
\begin{equation}
 v_0(\pt)=
 \frac{\left\langle\delta n(\pt)\,\delta \left[\pt\right]^{\rm ref}\right\rangle}
 {\left\langle n(\pt)\right\rangle\,\sigma_{\left[\pt\right]^{\rm ref}}},
 \qquad
 n(\pt)=\frac{1}{N_{\rm norm}}\frac{dN}{d\pt}.
 \label{eq:v0}
\end{equation}
This measures the event-by-event correlation between mean $\pt$ and multiplicity in each $\pt$ bin. Let us assume particle productions from hydrodynamics. If an event has larger $\left[\pt\right]$ than average, typically, it means the $p_T$ spectra from that event is harder.
This leads to $\left\langle\delta n(\pt)\,\delta \left[\pt\right]^{\rm ref}\right\rangle > 0$ at high $\pt$ while $\left\langle\delta n(\pt)\,\delta \left[\pt\right]^{\rm ref}\right\rangle<0$ at low $\pt$, producing a sign change in $v_0(\pt)$ around the characteristic momentum where the normalized spectrum pivots. 
This monotonic increase of $v_0(\pt)$ has been shown by several hydrodynamic simulations~\cite{Gardim:2019iah,Schenke:2020uqq,Du:2025dpu}. It is expected to provide information on the event-by-event fluctuation of radial flow, which certainly adds more information compared to the event-average observables.

Several new results on $v_0(\pt)$ were presented at this conference~\cite{YKong:SQM2026,ZWang:SQM2026,ADimri:SQM2026}.
Measurements with particle-identification can give more detailed insight into the mass-dependence of radial flow fluctuation. A theoretical model calculation showed that the heavy-flavor $v_0$ was sensitive to the interaction strength between heavy quarks and the medium \cite{SPlumari:SQM2026}. The measurement of $v_0(\pt)$ was also extended to the RHIC collision energies. 

At higher $\pt$-regime, $v_0(\pt)$ can show additional structure due to the interaction between soft and hard. 
After the monotonic increase at low $p_T$, the observed $v_0(p_T)$ starts decreasing at a few GeV, indicating that different dynamics than hydrodynamics governs the particle production at that regime~\cite{Du:2025hrz}.
This makes $v_0(\pt)$ an informative observable also for such soft--hard interaction. 
A major theoretical challenge is therefore to describe $v_0(\pt)$
consistently across the momentum region where soft and hard
particle production overlap.

\section{Summary}

In this contribution, I reviewed recent developments in the study of bulk QCD matter and light/strange hadron production. Light-ion collisions provide an important bridge between heavy-ion and small-system physics. Recent $OO$ results generally favor descriptions with hydrodynamics, while strangeness production and results from core--corona models indicate that non-equilibrium contributions may be sizable. Quantifying the applicability of hydrodynamics in such intermediate-size systems therefore remains an important open question.
Quantitative constraints on QGP properties require both improved initial-state modeling and new observables. Three-dimensional and more physics-driven initial conditions offer a path toward reducing model uncertainties. Radial flow fluctuation provides further information of event-by-event dynamics, and potentially soft--hard interactions at intermediate to high transverse momentum region. 

\vspace{10pt}
\noindent {\bf{Acknowledgment}} \quad The author would like to thank Volker Koch, Lipei Du, and Iurii Karpenko for useful discussions. The author is supported by the Japan Science and Technology Agency (JST) as part of Adopting Sustainable Partnerships for Innovative Research Ecosystem (ASPIRE) Grant No. JPMJAP2318.










\bibliographystyle{elsarticle-num}
\bibliography{sqm2026_template}



\end{document}